\documentclass[twocolumn,secnumarabic,amssymb, nobibnotes, aps, prd]{revtex4}
\usepackage{graphicx}

\begin{document}
\title{Quantum oscillations of the rectified voltage and the critical current of asymmetric mesoscopic superconducting loops}
\author{V.L. Gurtovoi , S.V. Dubonos, A.V. Nikulov, N.N. Osipov, and V.A. Tulin}
\affiliation{Institute of Microelectronics Technology and High Purity Materials, Russian Academy of Sciences, 142432 Chernogolovka, Moscow District, RUSSIA.} 
\begin{abstract}The current-voltage curves and magnetic dependence of the critical current of asymmetric superconducting loops are measured. It was found that sign and value of the asymmetry of the current-voltage curves changes with value of magnetic field, periodically for single loop and system of identical loops. The obtained results allow to explain the quantum oscillation of the dc voltage, observed below superconducting transition in the previous works, as rectification of ac current or noise.
 \end{abstract}

\maketitle

\narrowtext

\section*{Introduction}

The Bohr's quantization $\oint_{l} dl p = n2\pi \hbar $ postulated for the explanation of the stable electron orbits in atom is the cause of  various quantum phenomena [1] on the mesoscopic level. One of the most wonderful phenomena is the persistent current $I_{p}$ observed in normal metal [2], semiconductor [3] and superconductor [4,5] mesoscopic loop $l$. Its value and sign vary periodically with value of magnetic flux $\Phi$ inside the loop since the canonical momentum $p =mv + qA$ and therefore permitted values of velocity circulation $\oint_{l} dl v = 2\pi \hbar (n -\Phi/\Phi_{0})$, where $\Phi_{0}= 2\pi \hbar/q$ is the flux quantum. For the first time the persistent current $I_{p}(\Phi/\Phi_{0}) \propto n -\Phi/\Phi_{0}$ was predicted [4] and observed [5] in superconductor structure. Later the like periodical dependence $I(\Phi/\Phi_{0})$ was predicted [6] and observed [2,3] in non-superconducting mesoscopic structures. Recently the quantum oscillations of the dc potential difference $V_{dc}(\Phi/\Phi_{0}) \propto I_{p}(\Phi/\Phi_{0})$ were observed on segments of asymmetric superconducting loops [7,8]. It is important to investigate the cause of  this phenomenon in order to clear up a question: "Could a like one be observed in normal metal or semiconductor asymmetric mesoscopic loops?"

It was found that the quantum oscillations $V_{dc}(\Phi/\Phi_{0})$ can be observed without an evident power source near superconducting transition [7], whereas at a lower temperature they can be induced by an external ac current when its amplitude exceeds a critical value [8]. It was assumed in [8] that the asymmetry of the current-voltage curves and its periodical change with magnetic field are the cause of the $V_{dc}(\Phi/\Phi_{0})$ observed in the both cases. In order to verify this assumption the current-voltage curves of asymmetric aluminum loops and its change with magnetic field are investigated in the present work.

\section {Experimental}

Investigated structures with thickness $d = 40-70 \ nm$ consisted of asymmetric aluminum round loops (rings) with semi-ring width $w_{n} = 200 \ nm$ and $w_{w} = 400 \ nm$ for narrow and wide parts, respectively, Fig.1. They were fabricated by thermally evaporated on oxidized Si substrates, e-beam lithography and lift-off process. Two single loops with diameter $d = 4 \ \mu m$, width of the current stripe, see Fig.1, $w_{con} = 0.6 \ \mu m$ and $w_{con} = 0.7 \ \mu m$, two systems of identical 20 loops with $d = 4 \ \mu m$, $w_{con} = 0.4 \ \mu m$ and $w_{con} = 1 \ \mu m$ and  two systems of double loops with different diameter $d = 4 \ \mu m$ and $d = 3 \ \mu m$ were investigated. For this structures, the sheet resistance was $0.2 \div 0.5 \ \Omega /\diamond $ at 4.2 K, the resistance ratio $R(300K)/R(4.2K)=2.5 \div 3.5$, and critical temperature was $T_{c} = 1.24 \div 1.27 \ K$.

The current-voltage curves and magnetic dependencies of the critical current $I_{c+}$, $I_{c-}$ measured in opposite directions on the loop, shown on Fig.1, and systems of such loops were investigated. Magnetic field direction was perpendicular to the ring's plane. All signals were digitized by a multi-channel 16-bit analog-digital converter card.

\begin{figure}[b]
\includegraphics{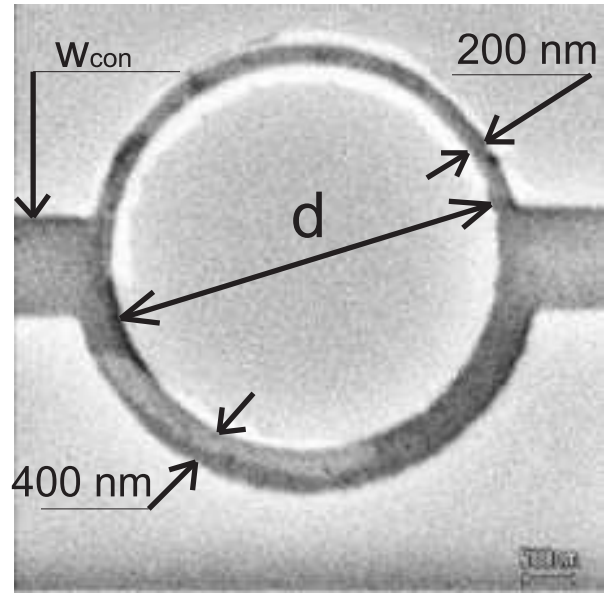}
\caption{\label{fig:epsart} An SEM photo of the asymmetric aluminum round loop (ring) used for the measurements. $w_{con}$ is the width of the Al stripes used as current contacts.}
\end{figure}

\section {Results}

Three types of the current-voltage curves are observed: 1) smooth and reversible one; 2) irreversible, with smooth transition into the resistive state; 3) irreversible, with jump increase of the resistance of the whole structure at $I = I_{c+}$ or $I_{c-}$. The first type is observed near superconducting transition $T_{c}$, the second one in an intermediate region of temperature and the third one at low temperature. We have found that the critical current $I_{c+}$, $I_{c-}$ both of single loops and systems of identical loops is periodical function of magnetic field with period corresponding to the flux quantum $\Phi_{0}$, Fig.2. The magnetic dependencies of the critical current $I_{c+}(\Phi/\Phi_{0})$ and $I_{c-}(\Phi/\Phi_{0})$ measured in opposite directions are similar, Fig.2, for all investigated structures. The anisotropy of the critical current $I_{c,an} = I_{c+} - I_{c-}$, Fig.2, is a consequance of a shift $\Delta \phi $ of these dependencies one relatively another, $I_{c+}(\Phi/\Phi_{0}) = I_{c-}(\Phi/\Phi_{0}+\Delta \phi )$.

\begin{figure}[b]
\includegraphics{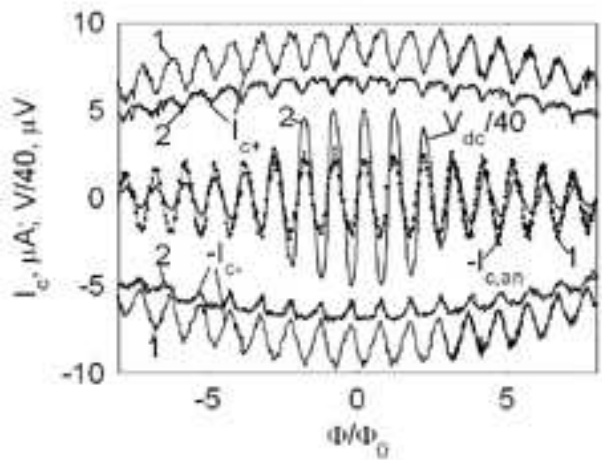}
\caption{\label{fig:epsart} The quantum oscillations of the critical current $I_{c+}$, $I_{c-}$ measured in opposite directions, its anisotropy $I_{c,an} = I_{c+}- I_{c-}$ and the rectified voltage $V_{dc}$, induced by the ac current with frequency f = 1 kHz and amplitude $I_{0} = 17.5 \ \mu A$ at T = 1.321 K. The data for single loop (1) and system of identical 20 loops (2) are shown.}
\end{figure}

The similarity of the magnetic periodical dependencies of anisotropy of the critical current $-I_{c,an}(\Phi/\Phi_{0})$  and the dc voltage $V_{dc}(\Phi/\Phi_{0})$ induced by an external ac current, Fig.2, corroborates the explanation [8] of the quantum oscillations of the dc voltage observed on segments of asymmetric superconducting loops as a consequence of rectification of the ac current [8] or noise [7]. The similarity of the magnetic dependencies $V_{dc}(B)$ and $-I_{c,an}(B)$ is observed also on double loops with different diameter, Fig.3, although these dependencies are not periodical in this case [9].

The rectified voltage $V_{dc} = \Theta ^{-1} \int_{\Theta }dtV(I_{ext}(t))$ appears when the amplitude $I_{0}$ of the external current $I_{ext}(t) = I_{0} \sin(2\pi ft)$ exceeds either  $I_{c+}$ or $I_{c-}$ and its absolute value $|V_{dc}|$ decreases when $I_{0}$ exceeds both $I_{c+}$ and $I_{c-}$. The amplitude $V_{A}$ of the quantum oscillations $V_{dc}(\Phi/\Phi_{0})$ has a maximum value [8,9] $V_{A,max}$ at $min(I_{c+},I_{c-}) < I_{0,max} < max(I_{c+},I_{c-})$. Our measurements have shown that the relation $V_{A,max}/I_{0,max}$ is high at a low temperature $T < 0.98T_{c}$ where the current-voltage curves of the third type are observed: $V_{A,max}/I_{0,max} \ \approx 0.8 \ \Omega $ for a single loop with the resistance in the normal state $R_{n} = 3.3 \ \Omega $ and $V_{A,max}/I_{0,max} \approx 20 \ \Omega $ for a systems of identical 20 loops with $R_{n} = 92 \ \Omega $. The high efficiency of rectification is conditioned by the irreversibility of the current-voltage curves and it decreases near superconducting transition.

The conclusion that the potential difference $V_{dc}(\Phi/\Phi_{0})$ observed at $T < T_{c}$ on asymmetric superconducting loop is result of rectification of an ac current or noise does not exclude a like phenomenon in normal metal or semiconductor asymmetric loops. The asymmetry of the current-voltage curves and $I_{c+}(\Phi/\Phi_{0})$, $I_{c-}(\Phi/\Phi_{0})$ are consequence of the persistent current $I_{p}(\Phi/\Phi_{0})$ which is observed not only at $T < T_{c}$ but also at $T \approx  T_{c}$ and even $T > T_{c}$ [5] where $R_{l} > 0$. One can expect $V(\Phi/\Phi_{0}) = R_{asym} I_{p}(\Phi/\Phi_{0})$ at $ I_{p}\neq 0$ and $R_{l} > 0$ by analogy with the case $V = R_{asym} I = (R_{ls} - R_{l}l_{s}/l)I$ of conventional circular current $I = R_{l}^{-1}\oint_{l}dlE$ induced by the Faraday's voltage $\oint_{l}dlE = -d\Phi/dt$ in an asymmetric loop with the segment resistance $R_{ls}/l_{s} \neq R_{l}/l$.

\begin{figure}
\includegraphics{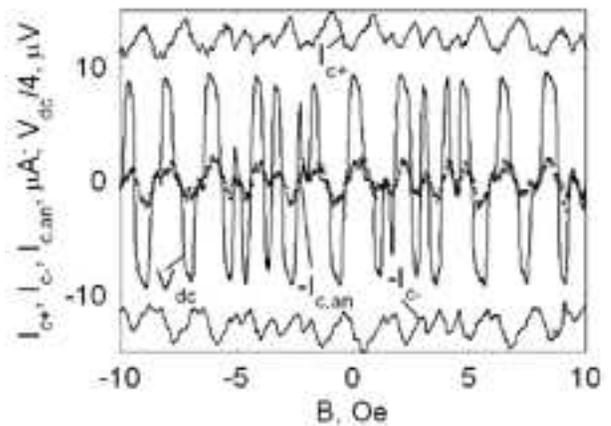}
\caption{\label{fig:epsart} The magnetic dependencies of the critical current $I_{c+}$, $I_{c-}$ measured in opposite directions, its anisotropy $I_{c,an} = I_{c+}- I_{c-}$ and the rectified voltage $V_{dc}$, induced by the ac current with frequency f = 1 kHz and amplitude $I_{0} = 17.5 \ \mu A$ at T = 1.321 K of double superconducting Al loops.}
\end{figure}

This work has been supported by a grant of the Program "Low-Dimensional Quantum Structures" of the Presidium of Russian Academy of Sciences, a grant "Quantum bit on base of micro- and nano-structures with metal conductivity" of the Program "Technology Basis of New Computing Methods" of ITCS department of RAS and a grant 04-02-17068 of the  Russian Foundation of Basic Research.


\begin{thebibliography}{99}

\bibitem{Tulin1} D.~V.~Nomokonov et al., {\em in Proceedings of 13th International Symposium "NANOSTRUCTURES: Physics and Technology"} St Petersburg: Ioffe Institute, 2005, p. 197; B.~Szafral and N.~M.~Peeters, {\em idid}, p. 203; O.~A.~Tkachenko et al., {\em idid}, p. 205

\bibitem{Tulin2} L.~P.~Levy et al., {\em Phys. Rev.Lett.} {\bf 64}, 2074 (1990); V.~Chandrasekhar et al., {\em idid} {\bf 67}, 3578 (1991); E.~M.~Q.~Jariwala et al., {\em idid} {\bf  86}, 1594 (2001).

\bibitem{Tulin3} D.~Mailly, C.~Chapelier, and A.~Benoit, {\em Phys.Rev.Lett.} {\bf 70}, 2020 (1993); B.~Reulet et al., {\em idid} {\bf 75}, 124 (1995); W.~Rabaud et al., {\em idid} {\bf 86}, 3124 (2001).

\bibitem{Tulin4} J.~M.~Blatt, {\em Phys.Rev.Lett.} {\bf 7}, 82 (1961).

\bibitem{Tulin5} W.~A.~Little and R.~D.~Parks, {\em Phys.Rev.Lett.} {\bf 9}, 9 (1962).

\bibitem{Tulin6} I.~O.~Kulik, {\em Pisma Zh.Eksp.Teor.Fiz.} {\bf 11}, 407 (1970) ({\em JETP Lett.} {\bf 11}, 275 (1970)).

\bibitem{Tulin7} S.~V.~Dubonos, V.~I.~Kuznetsov and A.~V.~Nikulov, {\em in Proceedings of 10th International Symposium "NANOSTRUCTURES: Physics and
Technology"} St Petersburg: Ioffe Institute, 2002, p. 350.

\bibitem{Tulin8} S.~V~.Dubonos et al.,  {\em Pisma Zh.Eksp.Teor.Fiz.} {\bf 77}, 439 (2003) ({\em JETP Lett.} {\bf 77}, 371 (2003)).

\bibitem{Tulin9} V.~L.~Gurtovoi , R.~V.~ Kholin, N.~N.~Osipov, and V.~A.~Tulin, {\em in Proceedings of 13th International Symposium "NANOSTRUCTURES: Physics and
Technology"} St Petersburg: Ioffe Institute, 2005, p. 211

\end{thebibliography}
\end{document}